\documentclass[aps,prb,twocolumn,showpacs]{revtex4}

\usepackage{epsfig}
\usepackage{graphicx}

\def \bmatA       {\mbox{\boldmath$\mathcal{A}$}}

\begin{document}

\title{Vacuum fluctuations and the spin current in mesoscopic
structures\\
with collinear magnetic order}
\author{Vitalii K.~Dugaev$^{1}$ and Patrick Bruno$^2$}
\affiliation{$^1$Department of Mathematics and Applied Physics,
Rzesz\'ow University of Technology, Al. Powsta\'nc\'ow Warszawy 6,
35-959 Rzesz\'ow, Poland\\
$^2$Max-Planck-Institut f\"ur Mikrostrukturphysik, Weinberg 2,
06120 Halle, Germany}
\date{\today }

\begin{abstract}
We show that in magnetic nanostructures with a {\it homogeneous}
magnetic order, the equilibrium spin current can be nonzero. For
example, this is the case of a wide magnetic ring with the
magnetization along the ring axis. The physical reason of this
effect is a variation of the orientation of anisotropy axis
inducing a spin torque acting on the magnetic ions. The mechanism
of the spin current generation is related to the quantum vacuum
fluctuations in the magnetic system.
\vskip0.5cm \noindent
\end{abstract}
\maketitle

One of the key problems in modern magnetoelectronics is related to
the creation and manipulation of the spin currents
\cite{prinz98,ohno99,ganichev02,hirsch99}.
It was shown recently that the spin current can be generated using
the flux of magnons in a nonequilibrium state of the magnetic
system \cite{meier03}. This idea was further developed for the
magnons in textured magnetic structures like magnetic rings in the
inhomogeneous external field \cite{schutz03,wang04}. In the latter
case, the existence of spin currents is related to a gauge field
for the motion of magnons \cite{dugaev05,bruno05}. The gauge field
and, correspondingly, the Berry phase for the adiabatic motion of
magnons appear naturally due to the nonvanishing chirality of the
magnetic ordering. In terms of the Berry phase, the magnon
mechanism of the spin current excitation can be better understood
for various types of magnetic structures \cite{dugaev05}.

In the case of magnetic systems with topological excitations (like
vortices or skyrmions) or a system subject to inhomogenous fields,
there appears an equilibrium spin current associated with the
transmission of angular momentum \cite{bruno05}. This current transfers
the torque acting on the disoriented magnetic moments.  The
equilibrium spin current is usually much stronger than the one
transferred by magnons because it does not require any spin wave
excitations. This mechanism of spin current generation can be
realized only in textured ferromagnets, and it disappears in case
of a homogeneous magnetization.

In real magnetic structures, one should also take into account a
possible inhomogeneity with respect to the orientation of
anisotropy axes. It induces a new type of a {\it topological}
Berry phase, which does not necessarily vanish when the usual
geometric phase is exactly zero \cite{bruno04}. An example of such
a system is a wide magnetic ring (thin-wall cylinder) with the
magnetization along the cylinder axis \cite{bruno04,dugaev05}. In
this system, the gauge field is related to the variation of
anisotropy axis (perpendicular to the cylinder surface).

We show below that the equilibrium spin current does not vanish in
the system with the homogeneous magnetization but inhomogeneous
anisotropy. This effect has purely quantum origin and is related
to the vacuum fluctuations in the magnetic system. The quantum
character of this effect makes it substantially different from the
earlier studied mechanism of the equilibrium spin current in
noncollinear ferromagnets \cite{katsura05,bruno05}, which can be
understood within a model of the classical moment.

The spin current appearing in a shaped mesoscopic structure (e.g.,
a nanoring) with the collinear magnetic ordering, transfers the
angular momentum. The resulting transferred torque acts on the
anisotropy axes, tending to arrange them homogeneously. In other
words, the spin current produces a mechanical stress in the
anisotropy-noncollinear system.

The physics of this phenomenon can be better understood if we
consider first a simple model with two quantum spins $S=1$ in the
magnetic field $B$ directed along the axis $z$, and with the
anisotropy axes oriented along ${\bf n}_1$ and ${\bf n}_2$ in the
$x-y$ plane for spins $S_1$ and $S_2$, respectively (Fig.~1). We
take ${\bf n}_1$ along $x$ and denote the angle between ${\bf
n}_1$ and ${\bf n}_2$ by $\phi $. The Hamiltonian is
\begin{equation}
\label{1}
H=-BS_{1z}+\frac{\lambda }2\left( {\bf n}_1\cdot {\bf S}_1\right) ^2-BS_{2z}
+\frac{\lambda }2\left( {\bf n}_2\cdot {\bf S}_2\right) ^2
-g{\bf S}_1\cdot {\bf S}_2,
\end{equation}
where $\lambda >0$ is the constant of anisotropy, and the last term
corresponds to the exchange interaction with the coupling constant
$g$. We assume $g>0$, which corresponds to the ferromagnetic
coupling. Obviously, in the ground state both spins are oriented
along $z$.

\begin{figure}
\begin{center}
\vspace*{-0.7cm}
\epsfxsize=6.5cm \epsfbox{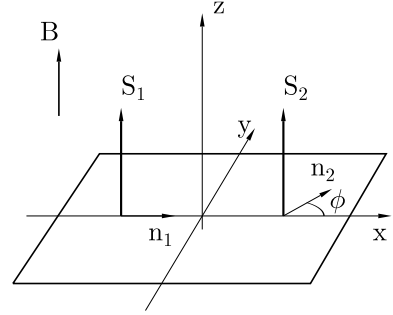}
\end{center}
\vspace*{-1.2cm}
\caption{Two spins $S=1$ with different orientation of anisotropy axes
in $x-y$ plane.}
\label{fig1}
\end{figure}

We can use the basis of eigenfunctions of Hamiltonian (1) with
$g=0$. For each of the noninteracting spins $S_1$ and $S_2$, the
eigenfunctions can be easily found, and they differ from each
other by the transformation $\exp (i\phi J_z)$, where $J_z$ is the
generator of $z$-rotations. In the absence of interaction, $g=0$,
the lowest four states of Hamiltonian (1) correspond to energies
$\left( 2\varepsilon _0,\; \varepsilon _0+\lambda /2,\; \lambda
\right) $, where the level $\varepsilon _0+\lambda /2$ is twofold
degenerate. We denote these states $\left| 00\right> , \left|
01\right> ,\left| 10\right> ,\left| 11\right> $. Here $\varepsilon
_0=\lambda /4-\left( \lambda ^2/16+B^2\right) ^{1/2}$ is the ground
state energy of a single spin.

The essential point is that each of two separated quantum spins in
the ground state has the magnitude of the moment along the $z$
axis smaller than unity, $\left| \left< S_{1z}\right> _0\right|
=\left| \left< S_{2z}\right> _0\right| =2\left| \varepsilon
_0\right| B/\left( \varepsilon _0^2+B^2\right) <1$. This is the
effect of the magnetic anisotropy. As a result, the quantum
fluctuations give a different value of the fluctuating moments
along the axes $x$ and $y$ (principal axes for $S_1$), $\left<
S_{1x}^2\right> _0-\left< S_{1y}^2\right> _0 =\left( \varepsilon
_0^2-B^2\right) /\left( \varepsilon _0^2+B^2\right) $. It leads to
the dependence of interaction on a mutual orientation of the $S_1$
and $S_2$ anisotropy axes.

The interaction $H_{int}=-g{\bf S}_1\cdot {\bf S}_2$ gives
nonzero matrix elements
\begin{eqnarray}
\label{2}
V\equiv \left< 00\right| H_{int}\left| 00\right>
=-\frac{4g\varepsilon _0^2B^2}{(\varepsilon _0^2+B^2)^2}\; ,
\end{eqnarray}
\begin{eqnarray}
\label{3}
P(\phi )\equiv \left< 00\right| H_{int}\left| 11\right>
=\left< 11\right| H_{int}\left| 00\right> ^*
\nonumber \\
=g\left( \cos \phi -\frac{2i\varepsilon _0B}
{\varepsilon _0^2+B^2}\; \sin \phi \right) .
\end{eqnarray}
Assuming $B\gg \lambda $, we can restrict ourselves by considering
only the four lowest energy states. Then the groundstate energy $E_0$
of the Hamiltonian (1) can be found by the diagonalization of
matrix $4\times 4$, and we obtain
\begin{equation}
\label{4}
E_0(\phi )=\tilde{\varepsilon }_0+\lambda /2
-\left[ \left( \tilde{\varepsilon }_0-\lambda /2\right) ^2
+\left| P(\phi )\right| ^2\right] ^{1/2},
\end{equation}
where $\tilde{\varepsilon }_0=\varepsilon _0+V/2$. Due to the
exchange interaction, the energy depends on the mutual orientation
of the anisotropy axis.
The torque acting on the spins is $\mathcal{M}(\phi )=-dE_0/d\phi
$. In the limit of $g\to 0$ we find from (4)
\begin{equation}
\label{5}
\mathcal{M}(\phi )
\simeq -\frac{g^2\left( \varepsilon _0^2-B^2\right) ^2\sin 2\phi }
{2\left( \varepsilon _0^2+B^2\right) ^2
\left( \lambda /2-\varepsilon _0\right) }\; .
\end{equation}
The torque vanishes for parallel or antiparallel configuration of
anisotropy axes, and also if they are perpendicular to each other,
$\phi =\pi /2$.

Let us consider now the model with arbitrary spins $S\ge 1$ on a
lattice, with a nearest-neighbor exchange interaction. We
assume a one-site anisotropy with the anisotropy axis ${\bf n}_i$
smoothly depending on position. The Hamiltonian has the following
form
\begin{equation}
\label{6}
H=-\frac{g}2 \sum _{<i,j>}{\bf S}_i\cdot {\bf S}_j
+\frac{\lambda }2\sum _i\left( {\bf n}_i\cdot {\bf S}_i\right) ^2.
\end{equation}
We assume that in the ground
state the ferromagnet has a uniform magnetization along the axis
$z$, and all vectors ${\bf n}_i$ lie in the $x-y$ plane. It
corresponds to the easy plane ferromagnet with a nonuniform
anisotropy and can be realized, for example, in a thin-wall
magnetic cylinder (Fig.~2) \cite{bruno04,dugaev05}.

Using a local frame with the $x$ axis along ${\bf n}_i$ in each
lattice site, one transforms (1) into
\begin{equation}
\label{7}
H=-\frac{g}2\sum _{<i,j>}S^T_i\, e^{i(\phi _i-\phi _j)J_z}S_j
+\frac{\lambda }2\sum _i (S^x_{i})^2,
\end{equation}
where $\phi _i$ is the rotation angle, and we use a matrix form of
presentation for the spin operators, $S^T_i=\left( S^x_i,\,
S^y_i,\, S^z_i\right) $.

After the Fourier transformation, $S_i^\mu =\sum _{\bf q}S_{\bf
q}^\mu \, e^{i{\bf q}\cdot {\bf r}_i}$, and taking the limit of
$q\to 0$ the Hamiltonian is
\begin{eqnarray}
\label{8}
H=\frac{a}2\sum _{\bf q}
\left[ S^T_{-{\bf q}}\, q^2\, S_{\bf q}
+2i({\bf q}\cdot \bmatA )
\left( S^x_{-{\bf q}}S^y_{\bf q}-S^y_{-{\bf q}}S^x_{\bf q}\right)
\right. \nonumber \\ \left.
+\bmatA ^2\left( S^x_{-{\bf q}}S^x_{\bf q}+S^y_{-{\bf q}}S^y_{\bf q}\right)
\right]
+\frac{\lambda }2\sum _i (S^x_{i})^2,
\end{eqnarray}
where $\bmatA =\nabla _i \phi _i$ is the gauge potential,
$a=ga_0^2$, $a_0$ is the
lattice constant, and we assume that the variation of $\bmatA $ in
space is small at the wavelength of magnons (adiabatic
approximation) \cite{dugaev05}.

We use the Holstein-Primakoff representation of spin operators
\cite{holstein40}, $S^z_i=S-n_i$, $S^+_i=\sqrt{2S}\left(
1-n_i/2S\right) ^{1/2}a_i$, $S^-_i=\sqrt{2S}\, a^+_i\left(
1-n_i/2S\right) ^{1/2}$, $n_i=a^+_ia_i$, and restrict ourselves by
the harmonic approximation in $a_i,\, a^+_i$. Here $a^+_i, a_i$
are the boson creation and annihilation operators for the magnons.
Then we find
\begin{eqnarray}
\label{9}
H=\frac{S}2 \sum _{\bf q} \left\{
\left[ \left( a\left( {\bf q}+\bmatA\right) ^2+\lambda _1\right)
\left( a^+_{\bf q}a_{\bf q}+a_{\bf q}a^+_{\bf q}\right) \right]
\right. \nonumber \\ \left.
+\lambda _2\left( a_{\bf q}a_{-{\bf q}}+a^+_{\bf q}a^+_{-{\bf q}}\right)
\right\}
\end{eqnarray}
where $\lambda_1=\lambda \left( 1-1/2S\right)$ and
$\lambda_2=\lambda \left( 1-1/2S\right) ^{1/2}$.

In the following, we make a change in Eq.~(9) substituting $\lambda _2\to
\lambda _1=\lambda \left( 1-1/2S\right) $. This is related to the
well-known renormalization of the constant $\lambda _2$ due to the
magnon interactions \cite{rastelli75}. It reproduces correctly
both $S\to 1/2$ and $S\to \infty $ limits, and leads to the
gapless spectrum of magnons in correspondence
with the Goldstone theorem \cite{kaganov88}.
The Hamiltonian (9) with $\lambda _2\to \lambda _1$ can be
diagonalized using the Bogolyubov-Holstein-Primakoff
transformation method \cite{holstein40}, and we obtain finally
\begin{equation}
\label{10}
H=\sum _{\bf q} \omega _{\bf q}\left( b^+_{\bf q}b_{\bf q}+1/2 \right) ,
\end{equation}
where $b^+_{\bf q},\, b_{\bf q}$ operators are related to
$a^+_{\bf q},\, a_{\bf q}$ by the Bogolyubov-Holstein-Primakoff
transformation
\begin{eqnarray}
\label{11}
b_{\bf q}=\alpha _{\bf q}a_{\bf q}+\beta _{\bf q}a^\dag _{-{\bf q}},\hskip0.3cm
b^\dag _{\bf q}=\alpha _{\bf q}^*a^\dag _{\bf q}+\beta _{\bf q}^*a_{-{\bf q}}
\end{eqnarray}
and
$\omega _{\bf q}$ is the energy spectrum of magnons,
\begin{equation}
\label{12}
\omega _{\bf q}=S\left( \tilde{J}_{\bf q}-\tilde{J}_{-{\bf q}}\right)
+S\left[ \left( \tilde{J}_{\bf q}+\tilde{J}_{-{\bf q}}\right)
\left( \tilde{J}_{\bf q}+\tilde{J}_{-{\bf q}}+2\lambda _1\right) \right] ^{1/2}.
\end{equation}
Here we denoted
\begin{equation}
\label{13}
\tilde{J}_{\bf q}\equiv J(0)-J({\bf q})=a\left( {\bf q}+\bmatA \right) ^2
\end{equation}
and used the standard notation $J({\bf q})$ for the Fourier transform of
exchange interaction.

Using (12),(13) we can find the dependence of momentum ${\bf q}$ on
$\bmatA $. It determines the Berry phase acquired by the spin
wave,
which moves in the gauge potential related to the varying anisotropy.
We find that in case of adiabatic motion (see details in Ref.~\cite{dugaev05}),
the Berry phase along a contour C is
$\gamma _B(C)=\oint _C\widetilde{\bmatA}({\bf r})\, d{\bf r}$, where
$\widetilde{\bmatA }=
\bmatA \left[ 1+S^2\lambda _1^2/\left( \omega ^0_{{\bf q}}\right) ^2\right] ^{-1/2}$
is the effective gauge potential acting on magnons and
$\omega ^0_{{\bf q}}=2S\left[
aq^2\left( aq^2+\lambda _1\right) \right] ^{1/2}$.
For $\lambda _1\to 0$, we get
$\widetilde{\bmatA }=\bmatA$ but for a large anisotropy, $\lambda
_1\gg \omega ^0_{{\bf q}}/S$, the effective potential
$\widetilde{\bmatA }\simeq \bmatA \, \omega ^0_{\bf q}/S\lambda _1$ is small
comparing to $\bmatA $. Thus, the Berry phase of
magnons is determined by $\widetilde{\bmatA }({\bf r})$. It generalizes the
result of Ref.~\cite{dugaev05} for arbitrary quantum
spins on the lattice.

We consider now the contribution to the total energy of the vacuum
fluctuations, $E_0=\sum _{\bf q}(\omega _{\bf q}/2)$,
corresponding to the second term in (10). The energy $E_0$ depends
on the gauge potential $\bmatA $ resulting from the inhomogeneous
anisotropy.
To calculate the spin current density ${\bf j}^s$ we add to
$\bmatA $ an additional fictitious potential $\bmatA ^\prime ({\bf
r})$, which we put zero after calculation. Thus, in the previous
definition of $J_{\bf q}$ we substitute $\bmatA \to \bmatA +\bmatA
^\prime $. Note that $\bmatA ^\prime $, like the original
potential $\bmatA$, is associated with the spin transformation --
rotation around the axis $z$, which corresponds to the spin
current density ${\bf j}^s$ with the spin polarization along $z$.

We use the definition $j^s_i({\bf r})=\gamma \left[ \delta
E_0/\delta \mathcal{A}^\prime _{i}({\bf r})\right] _{\bmatA
^\prime =0}$ following from the connection between the spin
current conservation and the invariance with respect to rotations
in the spin space \cite{dugaev05}, where $\gamma $ is the
gyromagnetic ratio. Then we obtain
\begin{equation}
\label{14}
{\bf j}^s=\frac{2a\gamma S}{\Omega }\sum _{\bf q}
\left\{ {\bf q}+\frac{\bmatA\left( \tilde{J}_{\bf q}
+\tilde{J}_{-{\bf q}}+\lambda _1\right) }
{\left[ \left( \tilde{J}_{\bf q}+\tilde{J}_{-{\bf q}}\right)
\left( \tilde{J}_{\bf q}+\tilde{J}_{-{\bf q}}+2\lambda _1\right) \right] ^{1/2}}
\right\} ,
\end{equation}
where $\Omega $ is the volume.
In the usual bulk or an infinite 2D system, the first term
vanishes due to the inversion symmetry ${\bf q}\to -{\bf q}$, and,
taking the limit of $\mathcal{A}\to 0$, we find
\begin{equation}
\label{15}
{\bf j}^s=\frac{a\gamma S\bmatA }{\Omega }
\sum _{\bf q}\frac{2aq^2+\lambda _1}
{\left[ aq^2\left( aq^2+\lambda _1\right) \right] ^{1/2}}.
\end{equation}
Here the sum over momentum runs up to $q_{max}\simeq \pi/a_0$. We
can estimate it for the 3D case as
\begin{eqnarray}
\label{16}
\frac1{\Omega }\sum _{\bf q}\frac{2aq^2+\lambda _1}
{\left[ aq^2\left( aq^2+\lambda _1\right) \right] ^{1/2}}
\simeq \frac1{3\pi ^2}
\left[ \left( q_{max}^2+\kappa ^2\right) ^{3/2}
-\kappa ^2\right]
\nonumber \\
-\frac{\kappa ^2}{2\pi ^2}
\left[ \left( q_{max}^2+\kappa ^2\right) ^{1/2}-\kappa \right] ,
\hskip0.5cm
\end{eqnarray}
where $\kappa =(\lambda _1/a)^{1/2}$.

As follows from (15), the spin current flows along the gradient of
the anisotropy axis variation. In terms of the continuous
anisotropy field, ${\bf n}({\bf r})$, it can be written as
$j^{s\mu }_i\sim \epsilon ^{\mu \nu \lambda}n^\nu (\partial
n^\lambda /\partial r_i)$, where index $\mu $ refers to the spin
polarization and $\epsilon ^{\mu\nu\lambda}$ is the unit
antisymmetric tensor. The spin current is responsible for the transmission
of angular momentum and produces the torque rotating the
anisotropy axis at each site to orient all of them in the same
direction.

In the case of magnetic rings, the first term in (14) can be
nonvanishing, which is related to the Berry phase acquired by a
moment moving along the ring \cite{bruno04}. As we assume the
magnetic ordering uniform, the Berry phase has a non-geometric
origin but is determined by the topology of the ring
\cite{bruno04}. In its turn, the nonvanishing Berry phase affects
the inversion symmetry with respect to the motion along the ring
in opposite directions.

\begin{figure}
\begin{center}
\vspace*{-0.7cm}
\epsfxsize=8cm \epsfbox{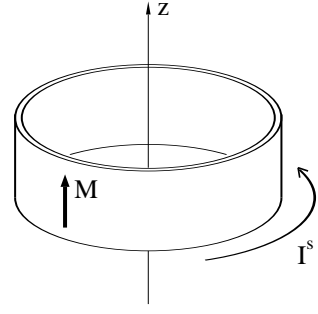}
\end{center}
\vspace*{-1.1cm}
\caption{Thin magnetic cylinder with the homogeneous magnetization and
noncollinear anisotropy.}
\label{fig2}
\end{figure}

To specify this suggestions, let us consider now the mesoscopic
magnetic ring in form of a thin-wall cylinder, with the
magnetization along the $z$ axis of cylinder, like described in
Ref.~\cite{dugaev05}. The hard axis is directed perpendicular to
the cylinder surface. Correspondingly, it changes direction along
a contour around the ring. We can restrict ourselves by
considering only the $\varphi $-component of momentum ${\bf q}$
corresponding to the around-ring motion (see Fig.~2)
because the gauge potential in this case has
only one nonvanishing component $\mathcal{A}_\varphi =1/R$, where $R$ is
the radius of the ring. Note that $\mathcal{A}_\varphi $ is a constant,
and it fully justifies using the Fourier transformation leading to
Eq.~(8). The periodic condition imposes the requirement
of momentum quantization, which reads as $\left(
q_{\varphi n}+\widetilde{\mathcal{A}}_\varphi \right) R=n$, where $n=0,\, \pm
1,...$. Then the first term in Eq.~(14) gives us the spin current
in the ring
\begin{equation}
\label{17}
\mathcal{I}^s=\frac{a\gamma S}{\pi R^2}\sum _n
\left[ n
-\left( 1+\frac{S^2\lambda _1^2}{\left( \omega ^0_{n}\right) ^2}\right) ^{-1/2}
\right] ,
\end{equation}
where $\omega ^0_{n}=2S\left[ \left( an^2/R^2\right) \left(
an^2/R^2+\lambda _1\right) \right] ^{1/2}$. In the limit of $\beta
\equiv \lambda _1R^2/a\gg 1$, which corresponds to the weak gauge
field regime, the main contribution is related to $n\sim
\sqrt{\beta }\gg 1$, and  we can substitute the sum in (17) by
the integral. In this approximation, we obtain
\begin{equation}
\label{18}
\mathcal{I}^s\simeq \frac{C\gamma S
\left( \lambda _1a\right) ^{1/2}}{\pi R},
\end{equation}
where
\begin{equation}
\label{19}
C=\int _{-\infty }^\infty dx
\left( 1-\frac{2|x|\sqrt{x^2+1}}{\sqrt{1+4x^2\left( x^2+1\right) }}
\right) \simeq 0.754.
\end{equation}
As follows from (18), the spin current is zero if the anisotropy
constant $\lambda _1=0$.
By evaluating the second contribution of Eq.~(14) into the spin
current, we find that in this case it is small with respect to
(18) in a parameter $a_0/(R\sqrt{\beta })\ll 1$.

Using (18) we can estimate the magnitude of the mechanical
stress in the ring. We take $\lambda _1=3.4\times
10^{-14}$~erg, corresponding to $\lambda _1/a_0^3=4\pi M^2$, $M=10^4$~G,
$a_0=3\times 10^{-8}$~cm; $a/a_0^2=4\times 10^{-14}$~erg, 
$2R=10$~nm. Then we get the angular momentum
$\mathcal{I}^a\simeq 3\times 10^{-16}$~erg. Varying the length of
ring $L$ by $\delta L$, we get a variation
$|\delta \mathcal{I}^a|=(\mathcal{I}^a/L)\, \delta L$.
Correspondingly the force acting in the cross section of the ring
is given by $f=\delta \mathcal{I}^a/\delta
L=\mathcal{I}^a/L$, and the stress $\sigma =f/S_0$, where $S_0$ is
the cross section of the ring. Taking $S_0=5\times
10^{-14}$~cm$^2$ ($5\times 1$~nm$^2$) we find $\sigma \simeq
0.05$~N/cm$^2$.

In conclusion, we demonstrated the presence of nonvanishing
equilibrium spin currents in magnetic structures with the
homogeneous magnetization and varied anisotropy axis. One of the
most important consequences of this effect is related to the
correct definition of the non-equilibrium spin current and
interpretation of spin current measurements
\cite{pzhang05,szhang05,sun05}. The existing attempts to exclude
the equilibrium spin current have been concentrated only to a part
responsible for the torque acting on spins. In the collinear
magnetic system, this part of torque is zero. Nevertheless, as we
demonstrated, the spin current can also transmit the angular
momentum rotating the anisotropy axes like in the case of magnetic
rings. Similar to the magneto-electric effect in magnetically
inhomogeneous systems \cite{bruno05}, the mechanical stress in
anisotropy-inhomogeneous systems can be called the {\it magneto-mechanical}
effect.

Our estimation of the stress suggests that the corresponding
deformations are too small to be measured experimentally.
However, the spin current in the ring can be observed due to the electric
polarization, as it was discussed recently 
\cite{meier03,schutz03,katsura05,bruno05}. 
It should be noted that the combination of magnetoelectric and magnetoelastic 
effects in the ferromagnet can be connected with the physics of ferroic 
materials, which are in scope of great activity now \cite{kimura03}.      

We calculated the spin current at $T=0$, when the excitation
of real magnons can be neglected. The effect of temperature is twofold.
Due to the excitation of magnons, one can expect an additional 
contribution to the spin current,
related to the Berry phase from the Dirac string \cite{bruno04}. 
On the other hand, the magnon decoherence increases with the temperature     
due to the magnon-phonon interaction, and it suppresses the spin current. 

This work is supported by FCT Grant POCI/FIS/58746/2004 in
Portugal, Polish Ministry of Science and Higher Education as a research
project in 2006--2009, and the STCU Grant No.~3098 in
Ukraine. V.D. thanks MPI f\"ur Mikrostrukturphysik in Halle for
the hospitality.

\end{document}